\documentstyle[12pt]{article}
\catcode`@=11

\def\address{\m@th\@ifnextchar[\@address{\@address[]}}

\def\@address[#1]#2{
\expandafter\def\expandafter\@addressname\expandafter
{\@addressname{
  \adr{#1}\ \parbox[t]{5in}{
     \ignorespaces #2}\par }}}
\def\@addressname{}
\def\adr#1{{\normalsize\unskip$^{#1}$}}
\def\@maketitle{%
\def\and{{\rm and}}
  \newpage
  \null
  {\centering
  \let \footnote \thanks
    {\Large\bf   \@title \par}%
    \vskip 1.5em%
      \lineskip .5em%
    {\bf\normalsize   \@author\par}
      \vspace{1em} 
    {\small \@addressname}
    
  }%
  \par
  \vskip 1.5em}
\def\section{\@startsection {section}{1}{\z@}{-3.5ex plus-1ex minus
    -.2ex}{1.5ex plus.2ex}{\reset@font\large\bf}}
\def\subsection{\@startsection{subsection}{2}{\z@}{-3.25ex plus-1ex
    minus-.2ex}{1.5ex plus.2ex}{\reset@font\normalsize\bf}}
\def\subsubsection{\@startsection
     {paragraph}{4}{\z@}{3.25ex plus1ex minus.2ex}{-1em}{\reset@font
     \normalsize\bf}}

\catcode`@=12

\title{Multibracket simple Lie algebras
\footnote{Talk at XXI Int. Coll. on Group Theor. Methods in 
Phys. (Goslar, July 1996), to appear in the Proceedings.}}
\author{J.A. de Azc\'{a}rraga\footnote{St. John's College Overseas Visiting 
Scholar.}
\hskip 0pt\ \and\ J.C. P\'{e}rez Bueno\footnote{
On sabbatical (J.A.) leave and on leave of absence (J.C.P.B.) 
from Departamento de F\'{\i}sica Te\'orica and IFIC
(Centro Mixto Univ. de Valencia-CSIC), E--46100 Burjassot (Valencia), Spain.}}
\address{Department of Applied Mathematics and Theoretical Physics, 
Silver St., Cambridge CB3 9EW, UK} 
\def\be{\begin{equation}}
\def\ee{\end{equation}}
\def\ie{{\it i.e.}}

\def\eq#1{(\ref{#1})}

\def\g{{\cal G}}
\begin{document}
\maketitle
\begin{abstract}
We introduce higher-order (or multibracket) simple Lie algebras that 
generalize the ordinary Lie algebras.
Their `structure constants' are given by 
Lie algebra cohomology cocycles which, by virtue of being such, satisfy a 
suitable generalization of the Jacobi identity.
Finally, we introduce a nilpotent, complete BRST operator associated with the 
$l$ multibracket algebras which are based on a given simple Lie algebra of 
rank $l$.
\end{abstract}
\noindent
Given $[X,Y]:=XY-YX$, the standard Jacobi identity (JI)
$[[X,Y],Z]+[[Y,Z],X]+[[Z,X],Y]=0$
is automatically satisfied if the product is associative.
For a Lie algebra $\g$,
$[X_i,X_j]=C_{ij}^k X_k$,
the JI may be written in terms of $C_{ij}^k$ as
\be
{1\over 2}\epsilon^{j_1j_2j_3}_{i_1i_2i_3}
C^\rho_{j_1 j_2}C^\sigma_{\rho j_3}=0\quad.
\label{JIb}
\ee
Let $\g$ be simple and (for simplicity) compact. 
Then, the Killing metric $k$, with coordinates $k_{ij}=k(X_i,X_j)$, is 
non-degenerate and, after suitable normalization, can be 
brought to the form $k_{ij}=\delta_{ij}$.
Moreover, $k$ is an invariant polynomial, \ie
\be
k([Y,X],Z)+k(X,[Y,Z])=0\quad.
\label{INV}
\ee
We also know that $k$ defines the second order Casimir invariant.
Using this symmetric polynomial we may always construct a non-trivial 
three-cocycle
\be
\omega_{i_1i_2i_3}:=k([X_{i_1},X_{i_2}],X_{i_3})=C_{i_1 i_2}^\rho k_{\rho i_3}
\label{threecocycle}
\ee
which is indeed skew-symmetric as consequence of \eq{JIb} or \eq{INV}.

In fact, it is known since the classical work of Cartan, Pontrjagin, Hopf and 
others that, from a topological point of view, the group manifolds of all 
simple compact groups are essentially equivalent to (have the [real] homology 
of) products of odd spheres, that $S^3$ is always present in these products and 
that the simple Lie algebra cocycles are, via the `localization' process, in 
one-to-one correspondence with bi-invariant de Rham cocycles on the associated 
compact group manifolds $G$.
This is due to the intimate relation between the order of the 
$l(=$rank$\,\g)$ 
primitive symmetric polynomials which can be defined on a simple Lie algebra, 
their $l$ associated generalized Casimir-Racah invariants \cite{RACAH} and the 
topology of the associated simple groups. Such a relation was a key fact in the 
eighties for the understanding of non-abelian anomalies in gauge theories 
\cite{TJZW}.

The simplest (of order 3)  higher-order invariant polynomial 
$d_{ijk}=d(X_i,X_j,X_k)$ appears for $su(3)$ (and only for $A_l$-type 
algebras, $l\ge 2$);
it is given by the symmetric trace of three $su(3)$ generators.
{}From $d_{ijk}$ we may construct
\be
\omega_{i_1i_2i_3i_4i_5}:=
\epsilon^{j_2 j_3 j_4}_{i_2 i_3 i_4} 
d([X_{i_1},X_{j_2}],[X_{j_3},X_{j_4}],X_{i_5})
=
\epsilon^{j_2 j_3 j_4}_{i_2 i_3 i_4} 
C_{i_1 j_2}^\rho C_{j_3 j_4}^\sigma d_{\rho \sigma i_5}
\label{fivecocycle}
\ee
(cf. \eq{threecocycle}), and it 
can be checked that \eq{fivecocycle} defines a fifth-order 
invariant form (the proof will be given in the general case).
The existence of this five-form $\omega$ shows us that $su(3)$ is, 
from a topological point of view, equivalent to $S^3\times S^5$.
If we calculate in $su(3)$ the `four-bracket' we find that
\be
[X_{j_1},X_{j_2},X_{j_3},X_{j_4}]=
\sum_{s\in S_4}\pi(s)X_{s(j_1)}X_{s(j_2)}X_{s(j_3)}X_{s(j_4)} 
={\omega_{j_1j_2j_3j_4}}^\sigma X_\sigma\quad,
\label{fourcomm}
\ee
where the generators $X_i$ may be taken proportional to the Gell-Mann matrices,
$X_i={\lambda_i \over 2}$, and $\pi(s)$ is the parity sign of the permutation 
$s$.
Thus, ${\omega_{j_1j_2j_3j_4}}^\sigma$ is related to the four-bracket and a 
five-cocycle (five-form) in the same way as ${C_{j_1j_2}}^\sigma$ is associated 
with the standard Lie bracket and a three-cocycle (three-form).

We may ask ourselves whether this construction could be extended to all the 
higher-order polynomials to define from them higher-order simple Lie algebras
satisfying an appropriate generalization of the JI.
The affirmative answer is given in \cite{HIGHER}; we outline below
the main steps that led to it.
It is interesting to note that this construction may be used to produce 
examples of a generalization \cite{APPB} of the Poisson structure different 
from that underlying Nambu mechanics \cite{Na}.

\medskip
\noindent
{\it a) Invariant polynomials on the Lie algebra $\g$}

Let $T_i$ be the elements of a representation of $\g$. Then the symmetric trace 
$k_{i_1\ldots i_m}\equiv\hbox{sTr}(T_{i_1}\ldots T_{i_m})$ (we shall only 
consider here sTr although not all invariant polynomials are of this form 
\cite{RACAH}; see \cite{HIGHER}) verifies the 
invariance condition 
\be
\sum_{s=1}^m C^{\rho}_{\nu i_s} k_{i_1\ldots i_{s-1} \rho i_{s+1}\ldots i_m}
=0\quad.
\label{invariance}
\ee
\medskip
\noindent
{\it Proof}: \quad 
By definition of $k$, the $l.h.s.$ of \eq{invariance} (cf. \eq{INV}) is 
\be
\hbox{sTr}\left(\sum_{s=1}^m T_{i_1}\ldots T_{i_{s-1}}[T_\nu,T_{i_s}]
T_{i_{s+1}}\ldots T_{i_m}\right) 
=
\hbox{sTr}\left(T_\nu T_{i_1}\ldots T_{i_m}- T_{i_1}\ldots T_{i_m}T_\nu 
\right)
= 0\ ,
\label{proof1}
\ee
{\it q.e.d.}
The above symmetric polynomial is associated to an invariant symmetric tensor 
field on the group $G$ associated with $\g$,
$k(g)=k_{i_1\ldots i_m}\omega^{i_1}(g)\otimes\ldots\otimes \omega^{i_m}(g)$, 
where the $\omega^{i}(g)$ are left invariant one-forms on $G$.
Since the Lie derivative of $\omega^k$ is given by 
$L_{X_i}\omega^k=-C^k_{ij}\omega ^j$ for a LI vector 
field $X_i$ on $G$, the invariance condition is the statement
\be
(L_{X_\nu} k)(X_{i_1},\ldots,X_{i_m})=
-\sum_{s=1}^m k(X_{i_1},\ldots,[X_\nu,X_{i_s}],\ldots,X_{i_m})=0
\label{liederivative}
\ee
{\it c.f.} \eq{INV}. 
On forms, the invariance condition \eq{liederivative} may be written as
\be
\epsilon^{j_1\ldots j_{q}}_{i_1\ldots i_{q}}C^{\rho}_{\nu j_1}
\omega_{\rho j_2\ldots j_q}=0\quad.
\label{formsinv}
\ee

\medskip
\noindent
{\it b) Invariant forms on the Lie group $G$}

Let
$k_{i_1\ldots i_m}$ be
an invariant symmetric polynomial on $\g$ and let us define 
\be
\tilde\omega_{\rho j_2\ldots j_{2m-2}\sigma}:=
k_{i_1\ldots i_{m-1}\sigma}
C^{i_1}_{\rho j_2}\ldots C^{i_{m-1}}_{j_{2m-3}j_{2m-2}}\quad.
\label{SNBa}
\ee
Then the odd order $(2m-1)$-tensor
\be
\omega_{\rho l_2\ldots l_{2m-2} \sigma}:=
\epsilon^{j_2\ldots j_{2m-2}}_{l_2\ldots l_{2m-2}}
\tilde\omega_{\rho j_2\ldots j_{2m-2} \sigma}
\label{SNBb}
\ee
is a fully skew-symmetric tensor. We refer to Lemma 8.1 in \cite{APPB} 
for the proof.

Moreover, $\omega$ is an invariant form because for $q=2m-1$ the 
$l.h.s.$ of \eq{formsinv} is
\begin{eqnarray}
&&\epsilon^{j_1\ldots j_{2m-1}}_{i_1\ldots i_{2m-1}}
C^{\rho}_{\nu j_1}\omega_{j_2\ldots j_{2m-1} \rho}
=
\epsilon^{j_1\ldots j_{2m-1}}_{i_1\ldots i_{2m-1}}
C^{\rho}_{\nu j_1}\epsilon^{l_3\ldots l_{2m-1}}_{j_3\ldots j_{2m-1}}
\tilde\omega_{j_2 l_3 \ldots l_{2m-1} \rho}
\nonumber \\
&&=
(2m-3)!\epsilon^{j_1\ldots j_{2m-1}}_{i_1\ldots i_{2m-1}}
k_{l_1\ldots l_{m}}C^{l_1}_{\nu j_1}\ldots C^{l_{m}}_{j_{2m-2}j_{2m-1}}
\label{proofb}
\\
&&=
(2m-3)!\epsilon^{j_1\ldots j_{2m-1}}_{i_1\ldots i_{2m-1}}
\left[\sum_{s=2}^m k_{\nu\l_2\ldots l_{s-1} \rho l_{s+1}\ldots l_{m}}
C^{\rho}_{j_1 l_s}\right] 
C^{l_2}_{j_2 j_3}\ldots  
C^{l_{m}}_{j_{2m-2}j_{2m-1}}=0\quad. 
\nonumber 
\end{eqnarray}
This result follows recalling
\be
\epsilon_{i_1\ldots i_p i_{p+1} \ldots i_n}^{j_1\ldots j_p j_{p+1} \ldots j_n}
\epsilon_{j_{p+1} \ldots j_n}^{l_{p+1} \ldots l_n}=
(n-p)!
\epsilon_{i_1\ldots i_p i_{p+1} \ldots i_n}^{j_1\ldots j_p l_{p+1} \ldots l_n}
\label{recordatorio}
\ee
in the second equality, using the invariance of $k$ [eq. \eq{invariance}] 
in the third one and the JI in the last equality for each of the $(m-1)$ terms 
in the bracket.

This may be seen without using coordinates; indeed \eq{SNBa} is expressed as 
\be
\tilde\omega(X_\rho,X_{j_2},\ldots,X_{j_{2m-2}},X_\sigma):=
k([X_\rho,X_{j_2}],\ldots,[X_{j_{2m-3}},X_{j_{2m-2}}],X_\sigma)
\label{newomega}
\quad,
\ee
and the ($2m$-1)-form $\omega$ is obtained antisymmetrizing \eq{newomega} 
as in \eq{SNBb}
(cf. \eq{fivecocycle}).
Hence
\begin{eqnarray}
&&
\hskip-25pt
(L_{X_\nu}\tilde\omega)(X_{i_1},\ldots,X_{i_{2m-1}})=
-\sum_{p=1}^{2m-1}
\tilde\omega(X_{i_1},\ldots,[X_\nu,X_{i_p}],\ldots,X_{i_{2m-1}})
\nonumber \\
&&
\hskip-25pt
=-\sum_{s=1}^{m-1}
k([X_{i_1},X_{i_2}],\ldots,[[X_\nu,X_{i_{2s-1}}],X_{i_{2s}}]+
[X_{i_{2s-1}},[X_\nu,X_{i_{2s}}]],\ldots,
\nonumber \\
&&
\hskip-25pt
[X_{i_{2m-3}},X_{i_{2m-2}}],X_{i_{2m-1}})
-k([X_{i_1},X_{i_2}],\ldots,[X_{i_{2m-3}},X_{i_{2m-2}}],[X_\nu,X_{i_{2m-1}}])
\nonumber \\
&&
\hskip-25pt
=-\sum_{s=1}^{m-1}
k([X_{i_1},X_{i_2}],\ldots,[X_\nu,[X_{i_{2s-1}},X_{i_{2s}}]],\ldots,
[X_{i_{2m-3}},X_{i_{2m-2}}],X_{i_{2m-1}})
\nonumber \\
&&
\hskip-25pt
-k([X_{i_1},X_{i_2}],\ldots,[X_{i_{2m-3}},X_{i_{2m-2}}],[X_\nu,X_{i_{2m-1}}])
\nonumber \\
&&
\hskip-25pt
=(L_{X_\nu} k)
([X_{i_1},X_{i_2}],\ldots,[X_{i_{2m-3}},X_{i_{2m-2}}],X_{i_{2m-1}})
=0
\quad;
\label{newproof}
\end{eqnarray}
where the JI has been used in the third equality
and \eq{liederivative} in the last, {\it q.e.d.}

\medskip
\noindent
{\it c) The generalized Jacobi condition}

Now we are ready to check that the tensor $\omega$ introduced above verifies a 
generalized Jacobi condition that extends eq. \eq{JIb} to multibracket 
algebras.

\medskip
\noindent
{\bf Theorem}\quad
Let $\g$ be a simple compact
algebra, and let $\omega$ be the
non-trivial Lie algebra $(2p+1)$-cocycle obtained from the 
associated $p$ invariant symmetric tensor on $\g$.
Then $\omega$ verifies the {\it generalized Jacobi condition} (GJC)
\be
\epsilon^{j_1\ldots j_{4p-1}}_{i_1\ldots i_{4p-1}}
{\omega_{\sigma j_1\ldots j_{2p-1}\cdot}}^\rho
{\omega_{\rho j_{2p} \ldots j_{4p-1}}}=0
\quad.
\label{theorem}
\ee
\noindent
{\it Proof:}\quad 
Using \eq{SNBb}, \eq{SNBa} and \eq{recordatorio}, the $l.h.s.$ of \eq{theorem} 
is equal to
\begin{eqnarray}
&&
-(2p-3)!\epsilon^{j_1\ldots j_{4p-1}}_{i_1\ldots i_{4p-1}}
k_{l_1 \ldots l_{p}\sigma} C^{l_1}_{\rho j_1} \ldots 
C^{l_{p}}_{j_{2p-2} j_{2p-1}}
{\omega^\rho_{\cdot j_{2p} \ldots j_{4p-1}}}
\nonumber 
\\
&&
=-(2p-3)!\epsilon^{j_1\ldots j_{4p-1}}_{i_1\ldots i_{4p-1}}
k^{l_1}_{\cdot \ldots l_{p}\sigma} C^{l_{2}}_{j_2 j_3}\ldots
C^{l_{p}}_{j_{2p-2} j_{2p-1}}C^\rho_{l_1 j_1}
\omega_{\rho j_{2p} \ldots j_{4p-1}} 
=0\quad,
\label{withname}
\end{eqnarray}
where the invariance of $\omega$ (eq. \eq{formsinv}) has been used in the last 
equality, {\it q.e.d.}

\medskip
\noindent
{\it d) Multibrackets and higher-order simple Lie algebras}

Eq. \eq{theorem} now allows us to define higher-order simple Lie algebras 
based on $\g$ using \cite{HIGHER} the Lie algebra cocycles $\omega$ on $\g$ as 
generalized structure constants:
\be
[X_{i_1},\ldots,X_{i_{2m-2}}]={\omega_{i_1\ldots i_{2m-2}}}^\sigma_\cdot 
X_\sigma
\quad.
\label{cocycle}
\ee
The GJC \eq{theorem} satisfied by the cocycles is necessary since for 
{\it even} $n$-brackets of associative operators one has the generalized Jacobi 
identity
\be
{1\over (n-1)!n!}\sum_{\sigma\in S_{2n-1}} (-)^{\pi(\sigma)}
[[X_{\sigma(1)},\ldots,X_{\sigma(n)}],X_{\sigma(n+1)},\ldots,X_{\sigma(2n-1)}]
=0\quad.
\label{genjacid}
\ee
This establishes the link between the $\g$-based {\it even} multibracket 
algebras and the {\it odd} Lie algebra cohomology cocycles on $\g$ 
(note that for $n$ odd the $l.h.s$ is proportional to the odd
($2n$-1)-multibracket $[X_1,\ldots,X_{2n-1}]$ \cite{HIGHER}).

Finally we comment that just in the same way that we can introduce for a Lie 
algebra a BRST nilpotent operator by
\be
s=-{1\over 2}c^ic^j{C_{ij}}^k_\cdot{\partial\over\partial c^k}\equiv s_2\quad,
\quad s^2=0\quad,
\label{BRST}
\ee
with $c^ic^j=-c^jc^i$, the set of invariant forms $\omega$ associated with a 
simple $\g$ allows us to 
{\it complete} this operator in the form
\begin{eqnarray}
s=
-{1\over 2}c^{j_1}c^{j_2}{\omega_{j_1j_2}}^\sigma_\cdot
{\partial\over\partial c^\sigma}
-\ldots-
{1\over (2m_i-2)!}c^{j_1}\ldots c^{j_{2m_i-2}}
{\omega_{j_1\ldots j_{2m_i-2}}}^\sigma_\cdot
{\partial\over\partial c^\sigma}
-\ldots 
\nonumber \\
-
{1\over (2m_l-2)!}c^{j_1}\ldots c^{j_{2m_l-2}}
{\omega_{j_1\ldots j_{2m_l-2}}}^\sigma_\cdot
{\partial\over\partial c^\sigma}
\equiv s_2+\ldots+s_{2m_i-2}+\ldots +s_{2m_l-2}.
\label{HOBRST}
\end{eqnarray}
This new nilpotent operator $s$ is the {\it complete BRST operator} 
\cite{HIGHER} associated with $\g$.

For the relation of these constructions with the strongly homotopy algebras 
\cite{LAST}, possible extensions and connections with physics we refer to 
\cite{HIGHER} and references therein.

\section*{Acknowledgements}
This research has been partially supported by the CICYT and DGICYT, Spain (AEN 
96--1669, PR 95--439). 
Both authors wish to thank the kind hospitality extended to them at 
DAMTP. Finally, the support of St. John's College (J.A.) and
an FPI grant from the Spanish M.E.C. and the CSIC
(J.C.P.B.) are gratefully acknowledged.


\end{document}